\documentstyle[emulateapj,epsfig,apjfonts]{article}

\newcommand\kms{\ifmmode {\rm \ km \ s^{-1}}\else $\rm km \ s^{-1}$\fi}
\newcommand\cm{\ifmmode {\rm \ cm }\else $\rm cm$\fi}
\newcommand\Mpc{\ifmmode {\rm \ Mpc }\else $\rm Mpc$\fi}
\newcommand\Myr{\ifmmode {\rm \ Myr }\else $\rm Myr$\fi}
\newcommand\kpc{\ifmmode {\rm \ kpc }\else $\rm kpc$\fi}
\newcommand\pc{\ifmmode {\rm \ pc }\else $\rm pc$\fi}
\newcommand\erg{\ifmmode {\rm \ erg }\else $\rm erg$\fi}
\newcommand\ergs{\ifmmode {\rm \ ergs }\else $\rm ergs$\fi}
\newcommand\s{\ifmmode {\rm \ s }\else $\rm s$\fi}
\newcommand\yr{\ifmmode {\rm \ yr }\else $\rm yr$\fi}
\newcommand\yrs{\ifmmode {\rm \ yrs }\else $\rm yrs$\fi}

\slugcomment{Submitted to ApJ Letters}

\lefthead{Ahn}
\righthead{Environment of GRB971214}

\begin{document}
\title{Environment of The Gamma-Ray Burst GRB971214 : A Giant H II Region
surrounded by A Galactic Supershell} 
\author{Sang-Hyeon Ahn}
\affil{Department of Astronomy, Seoul National University \\
San 56-1 Shillim-dong Kwanak-gu, Seoul, Korea}
\authoremail{sha@astro.snu.ac.kr}

\begin{abstract}
Among a number of gamma ray bursts whose host galaxies are known, GRB971214
stands out for its high redshift $z\ge 3$ and the Ly$\alpha$ emission line
having a P-Cygni type profile, which is interpreted
to be a direct consequence of the expanding supershell. From a profile
fitting analysis we estimate the expansion velocity of the supershell
$v_{exp} = 1500\kms$ and the neutral column density
$N_{HI}=10^{20}\cm^{-2}$. The redshift $z=3.418$ of
the host galaxy proposed by Kulkarni et al. (1998) has been revised 
to be $z=3.425$ from our profile analysis.

The observed Ly$\alpha$ profile is fitted well by a Gaussian curve,
which yields the Ly$\alpha$ luminosity
$L_{Ly\alpha}=(1.8\pm0.8)\times10^{42}\ergs \s^{-1}$.
Assuming that the photon source is a giant H II region,
we deduce the electron number density in the H II region
$n_e = (40\pm10) ({ L \over L_{Ly\alpha}})^{0.5} 
({R \over {100 \pc}})^{-1.5}\ \cm^{-3}$, which corresponds
to the illumination by about $10^4$ O5 stars.
We estimate the star-formation rate to be
$R_{SF} = (7\pm3){\rm M}_\odot\yr^{-1}$
with the internal and the Galactic extinction corrected.

The theory on the evolution of supernova remnants is used to propose
that the supershell is at the adiabatic phase, with its radius
$R = 18\ E_{53}^{1/2}\ \pc$, its age
$t = 4.7\times10^3\ E_{53}^{1/2} \yrs$, and the density of the ambient
medium $n_1 = 5.4\ E_{53}^{-1/2}\cm^{-3}$,
where $E_{53}= E/10^{53}\ergs$.
And we estimate the kinetic energy of the supershell to be
$E_k=7.3\times10^{52}\ E_{53} \ergs$.
These values are consistent with the hypothesis
that the supershell is the remnant of a gamma ray burst.

We note similarities between supershells found in nearby galaxies 
and remote primeval galaxies, and propose that the gamma ray burst may 
have occured in a giant H II region whose environment is similar to 
in star forming galaxies.

\end{abstract}
\section{Introduction}
 
Gamma ray bursts (hereafter GRBs), located at cosmological distances, 
form a group of the most luminous objects in the Universe. A number of GRBs were observed with their host galaxies, of which
intensive spectroscopic and imaging observations have been performed
using the Hubble Space Telescope and the Keck telescopes.
GRB971214 is one of such objects with a very high redshift $z > 3$ and also 
is worth a particular attention in following points.

Firstly, even after the optical transient region had faded away, 
we can marginally detect a bright spot in the HST image,
and the continuum and emission line fluxes from this spot overwhelm 
those from the remaining part of the host galaxy.
Secondly, the ultraviolet spectrum of GRB971214 illustrated 
in Fig.~1 shows a flat UV continuum which is often found in star forming 
galaxies. In addition, the Ly$\alpha$ emission line has a black absorption 
trough in the red part of the emission peak. These facts imply that the UV
spectrum is formed in a star forming region which is surrounded by
a thick and expanding medium of neutral hydrogens. We note that the location 
of the GRB afterglow coincides with the star forming region or
the bright spot, which leads to the proposal that the GRB occurs 
in a star forming region, and in this Letter we deduce 
the physical environment of the GRB from its spectrum.

\begin{figure*}[t]
\epsscale{0.7}
\plotone{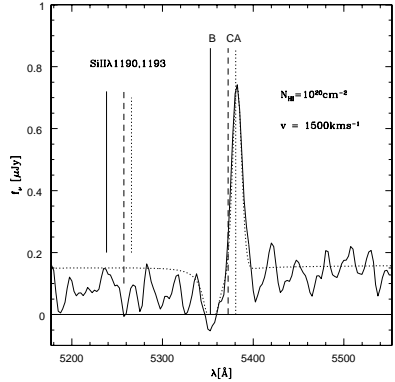}
\caption{\footnotesize
Profile of the Ly$\alpha$ emission line
and its P-Cygni type absorption (solid line) with the best fit profile
(dotted line). The emission has a Gaussian
profile whose width is $5.0\ {\rm \AA}$,
line center flux $f(\lambda=5380.8\ {\rm \AA})=0.675\ \mu$Jy.
The line center frequency $\lambda=5380.8\ {\rm \AA}$ gives
$z=3.425$. The P-Cygni type absorption is best fitted
by the column density of the supershell $N_{HI}=10^{20}\cm^{-2}$
and the expansion velocity $v_{exp}=1500\kms$. The dotted vertical line
denoted by 'A' is the solid line center of our result, the line
denoted by B is the location of the P-Cygni absorption trough,
and the 'C' vertical line is provided by Kulkarni et al. (1998).
Comparing with Ly$\alpha$ we also show SiII$\lambda$1190 line, but
the S/N ratio is not good.
}
\end{figure*}

The P-Cygni type Ly$\alpha$ emission in the spectrum of primeval
galaxies has been often attributed to an absorption effect 
by a galaxy not associated with the primeval galaxy 
but intervening accidentally in the line of sight. 
It has been regarded as a damped Ly$\alpha$ absorption
that occurs in the vicinity of the source galaxy. 
In order to check this possibility, we calculate the probability 
for observing an intervening galaxy in front of the GRB host galaxy,
which is none other than the optical depth 
for seeing a galaxy between the GRB host galaxy at $z=3.425$
and the place that corresponds to $v_{exp}$ in the Hubble's expansion law.
The optical depth is simply expressed by 
\begin{equation}
\tau=n_g(1+z)^3 \sigma L,
\end{equation} 
where the comoving volume number density of normal galaxies at $z=0$,
$n_g\sim0.02h^3\Mpc^{-3}$ (Im 1995), and the path length 
$L$ is estimated to be $L=v_{exp}/H$ with $H$ being 
the Hubble constant at the redshift and given by
\begin{equation}
H=H_0 [\Omega_M (1+z)^3 + \Omega_{\Lambda} ]^{1/2},
\end{equation}
where the cosmological density parameters are 
$\Omega_M = 8\pi G \rho_0 / 3H_0^2 $,
$\Omega_{\Lambda} = \Lambda / 3H_0^2 $, and the Hubble parameter 
$H_0=65\kms\Mpc^{-1}$.  
Here, the cross section is given by $\sigma=\pi (r_{10} 10h^{-1}\kpc)^2$,
where $r_{10}$ is the typical galaxy size in units of $10\kpc$. 
A direct substitution yields the optical depth $\tau = 0.0023 r_{10}^2$
for $\Omega_M=1/3$ and $\Omega_{\Lambda}=2/3$,
and $\tau = 0.0025 r_{10}^2$ for 
$\Omega_M=0.3$ and $\Omega_{\Lambda}=0$.
This indicates that if the average size of galaxies at $z\approx 3$
is not large, then it is highly improbable that the damped absorption 
in the spectrum of host galaxy of GRB971214 is formed 
by a galaxy intervening accidentally. 

This leaves us to consider an alternative hypothesis, according to which 
the P-Cygni type profile of Ly$\alpha$ is formed by the the expanding 
supershell that surrounds the star forming region in GRB971214 and
is the remnant of the GRB precedent to GRB971214. 
In order to check this possibility, we now calculate a number of GRB event
in a galaxy at $z\simeq 3.4$. 
Assuming the supernova rate is proportional to the starforming rate,
Sadat et al. (1998) calculated the supernova rate at $z\approx 3.4$,
$\Gamma_{SN}\simeq 0.011\times 10^6\ h_{65}^3\ {\rm SNe} \Myr^{-1} \Mpc^{-3}$,
where $H_0=65\ h_{65} \kms \Mpc^{-1}$.
Accepting the concept that the GRB rate traces the massive star 
formation rate, Woods \& Loeb (1998) showed 
$\Gamma_{GRB} \simeq 10^{-6} \Gamma_{SN}$,
where $\Omega_M=0.3$, $\Omega_{\Lambda}=0.7$, 
and $H_0=65\ h_{65} \kms\Mpc^{-1}$.  Therefore, 
$\Gamma_{GRB} (z\approx 3.4) \simeq 0.011\ h_{65}^3 \Myr^{-1}\Mpc^{-3}$.
Adopting the number density of galaxies at $2.0<z<3.5$, 
$\Phi^* = 1.76\times 10^{-3}\ h_{65}^3 \Mpc^{-3}$ (Pozzetti et al. 1998),
we can get the GRB rate per a galaxy at $z\approx3.4$,
$\Gamma_{GRB} \simeq 6 \Myr^{-1}$.
Therefore, the number of GRB events per a galaxy during $10^{4-5}\yrs$ 
is $N_{GRB}=0.06 \sim 0.6$. Moreover, the beaming factor, if exists, 
can increase the event rate by another factor of ten and 
the number of GRB events per a galaxy during $10^{4-5}\yrs$
is $N_{GRB}=0.6 \sim 6$, which makes our supershell hypothesis 
more probable and alternative suggestion.

It is noticeable that the similar P-Cygni features are observed in 
primeval galaxies and nearby star forming galaxies. 
Lee and Ahn (1998) proposed that the features might be caused 
not by an overlapping intergalactic medium but by the expanding 
medium enveloping the star forming region in the galaxies.
And this concept may be applied to such a remote star-forming galaxy 
as the host galaxy of GRB971214. However, we can not exclude 
the possibility that the multiple supernovae explosion may also 
result in the expanding shell.

In this Letter, we adopt the hypothesis of GRB-driven supershell 
and perform a profile fitting analysis to derive physical parameters 
characterizing the expanding supershell of neutral hydrogens 
surrounding the star forming region of the GRB host galaxy.

\section{Images and Spectrum of GRB971214}

GRB971214 was detected at 9 UT, December 14, 1997 (Heise et al. 1997),
and its optical counterpart twelve hours after the burst (Halpern et al. 1998).
With a total fluence of $1.09 \times 10^{-5} \erg$s$ \cm^{-2}$ 
(Kippen et al. 1997) and the measured redshift
of $z=3.418$ (Kulkarni et al. 1998), its energy release is estimated to be
$\sim 3\times10^{53} \erg$s in $\gamma$-rays alone under the assumption
of isotropic emission, $\Omega_0=0.3$, and $H_0 = 65 \kms\Mpc^{-1}$. 

\begin{figure*}[b]
\epsscale{0.7}
\plotone{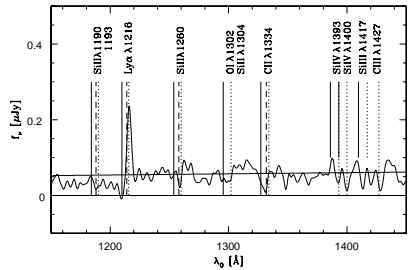}
\caption{\footnotesize
UV spectrum of the GRB971214 which is smoothed by
a Gaussian kernel with its width $\sigma=5\ {\rm \AA}$. Solid horizontal
line stands for the continuum level, the solid vertical bars represent
the line centers whose velocity components correspond to
the Ly$\alpha$ absorption. The refined absorption centers
which correspond to the systemic redshift of the GRB host
are presented by the dotted vertical lines. The dashed vertical bars
stand for the absorption lines suggested by Kulkarni et al. (1998)
The transitions for the absorption are also denoted. 
\label{fig2a}}
\end{figure*}

In Fig. 1 is shown its ultraviolet spectrum redshifted to the optical
band and obtained by Kulkarni et al. (1998) using the Keck telescope. It is 
characterized by a flat UV continuum that is typically found in the spectra 
of star forming regions.
It is also seen that the Ly$\alpha$ emission has a P-Cygni type profile,
which is frequently observed in the astronomical objects near or far
(Lee \& Ahn 1998). Lee and Ahn (1998) proposed that 
the P-Cygni type Ly$\alpha$ line is formed when the Ly$\alpha$ photons 
emitted in the central super star-cluster are radiatively transferred 
in a HI supershell that are optically thick and expanding.

In this work, the photon source is assumed to be
the H II region that may contain $10^4$ O stars.
According to Marlowe et al.(1995), nearby starbursting dwarfs
are inferred to contain a similar number of OB stars,
when considering H$\alpha$ luminosity. So this can be thought to be
neither entirely new nor extreme assumption for galaxies
of higher redshifts.
Furthermore, it appears less plausible that the Ly$\alpha$ emission arises
from the medium ionized by shocks produced by supernovae or hypernovae.
This is because the number density of the inner region
is not sufficiently high to give the recombination time scale
$\le 10^5$ years.

\section{Is the Photon Source Surrounded by the GRB Remnant? }
\subsection{Interpretation of the P-Cygni Absorption}

In this work we will consider the shell hypothesis,
and derive the physical properties of the expanding neutral medium 
from the observed Ly$\alpha$ absorption.

We assume a Gaussian profile for the unobscured Ly$\alpha$ emission and 
convolve it with a Voigt function with the center displaced by the expanding 
velocity that will be determined by the fitting procedure.
In principle, the effect of the frequency redistribution by back-scatterings 
should be considered. However, we neglect this effect in this paper, 
because the S/N ratio and the resolution of the spectrum are not 
sufficiently good.
For the continuum level, 
the blue part of Ly$\alpha$, which is more prone to extinction,
is extrapolated from the red portion of the spectrum given by 
Kulkarni et al. (1998). They quote
$F_{\nu}=174(\nu/\nu_R)^{\alpha}$ nJy with $\alpha=-0.7\pm0.2$,
where $F_{\nu}$ is the spectral density at frequency $\nu$ and
$\nu_R = 4.7\times10^{14}$ Hz, the central frequency of the R band.

\begin{figure*}[b]
\epsscale{0.6}
\plotone{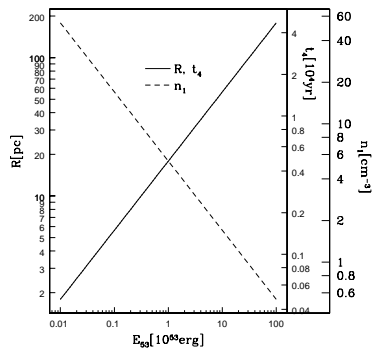}
\caption{\footnotesize
Radius($R$) in $\pc$, age($t_4$) in $10^4\yrs$,
and the volume number density of the ambient medium($n_1$)
with the variation of the input energy ($E_{53}$) in $10^{53}\ergs$.
\label{fig3}}
\end{figure*}

We show the result in Fig.~1, where the dotted line represents the best fit 
profile, the solid line the observed profile, and the horizontal solid line 
the continuum level.
The best fit expansion velocity of the supershell relative to the H~II region 
is determined to be $v_{exp} = 1500\kms$, 
and the best fit line center optical depth 
$\tau_0=6\times10^6$, which corresponds to $N_{HI}=10^{20}\cm^{-2}$.
The best fit Ly$\alpha$ profile has
the width of $\sigma=5\ {\rm \AA}$ and the line center flux 
$f(\lambda=5280.8\ {\rm \AA})=0.675\ \mu$Jy, which gives 
the unobscured flux to be $9.1\times10^{-18}\erg\cm^{-2}\s^{-1}$ and
the systemic redshift $z=3.425$.

This is slightly larger than the redshift proposed 
by Kulkarni et al. (1998), who may have overestimated the absorption 
in the blue part of the Ly$\alpha$. However, the absorption trough 
is sufficiently remote from the line center in the velocity space 
only to erode the extreme blue part of the Ly$\alpha$ emission. 
Hence, we prefer the redshift of $z=3.425$ of GRB971214
to the redshift of $z=3.418$, and 
subsequently other physical parameters need to be revised.

Assuming a standard Friedman cosmology with
$H_0=65\kms\Mpc^{-1}$ and $\Omega_0=0.3$, the luminosity
distance $d_L = 9.7\times10^{28}\cm$.
Considering the Galactic extinction, the unobscured Ly$\alpha$ flux is
corrected to be $F_{Ly\alpha} = (1.5\pm0.7)\times10^{-17}\erg\cm^{-2}\s^{-1}$,
where the observational error given by Kulkarni et al. (1998) is introduced.
Therefore, for the assumed cosmology, the Ly$\alpha$ line luminosity 
$L_{Ly\alpha} = (1.8\pm0.8)\times 10^{42} \erg \s^{-1}$.
If there is no internal extinction in the interior of the Ly$\alpha$ source, 
this corresponds to 
$n_e = (1.4\pm0.4) ({ L \over L_{Ly\alpha}})^{0.5}
({R \over {1 \kpc}})^{-1.5}\ \cm^{-3}$ or 
$n_e = (40\pm10) ({ L \over L_{Ly\alpha}})^{0.5}
({R \over {100 \pc}})^{-1.5}\ \cm^{-3}$, of which the ionization can be maintained
by $\sim 10^4$ O5 stars as the ionizing source. 

From the Ly$\alpha$ luminosity,
we can estimate the star-formation rate 
(Thompson, Djorgovski, \& Trauger 1995)
to be $R_{SF} = (7\pm3){\rm M}_\odot\yr^{-1}$,
with both the internal and the Galactic extinction being corrected.
This is consistent with the star forming rate
given by Kulkarni et al. (1998) as a lower limit,
$R_{SF}=5.2\ {\rm M}_\odot\yr^{-1}$
which was obtained from the rest-frame continuum luminosity 
at $1,500\ {\rm \AA}$.

Using the revised redshift of the GRB host galaxy, 
we refine other absorption lines in the observed spectrum.
In Fig.~2, we show the spectrum of the GRB host in the UV regime.
It is seen that the revised
wavelengths are in good agreement with the absorption features.

\subsection{Physical Configuration of the Supershell}

We consider the dynamical 
evolutionary model of the supernova remnant
to derive the physical quantities of the shell.
According to Woltjer (1972), the supernova remnant
has four evolutionary phases, that is, the free expansion 
phase, the Sedov-Taylor or adiabatic phase, the snowplow or radiative phase,
and finally the merging or dissipation phase 
(see also Reynolds 1988).

According to Woltjer(1972), the radiative phase begins
roughly when the expansion velocity of the shell becomes
\begin{equation}
v = 300\ ({n_1\over 1\cm^{-3}})^{2/17}({E \over 10^{53} \ergs})^{1/17}\ \kms,
\end{equation}
where $n_1$ is the number density of the ambient medium and
$E$ is the initial explosion energy.
Since the expansion velocity $v=v_{exp}=1500\kms$ of the supershell 
exceeds the velocity in the radiative phase by a large margin, we propose 
that the supershell is in the adiabatic phase, which is described by the 
Sedov solution.

According to the Sedov solution in a uniform medium of 
number density $n_1$ in which we have the relation $n_1 = 3N/R$,

\begin{equation}
R = 0.92\ ( { E \over m_H N })^{1/4} t^{1/2} ~\cm,
\end{equation}

\begin{equation}
v = 0.37\ ( { E \over m_H N } )^{1/4} t^{-1/2} ~\cm\s^{-1}, 
\end{equation}

\begin{equation}
n_1 = 3.2\ ({m_H N^5 \over E})^{1/4} t^{-1/2} ~\cm^{-3},
\end{equation}
where $v$ is the expansion velocity of the supershell,
$R$ the size of the supershell, $E$ the initial explosion
energy, $N$ the column density of the supershell,
$m_H$ the hydrogen mass, and $t$ the age of the shell.

Using the values $v = v_{exp} = 1500 \kms$ and $N=10^{20}\cm^{-2}$, we get

\begin{equation}
t = 4.7\times10^3\ ( {E_{53} \over N_{20}} )^{1/2}\ \yrs,
\end{equation}

\begin{equation}
R = 18\ ( { E_{53} \over N_{20}} )^{1/2}\ \pc,
\end{equation}

\begin{equation}
n_1 = 5.4\ ( {N_{20}^3 \over E_{53}} )^{1/2}\ \cm^{-3},
\end{equation}
where $N_{20}=N/10^{20}\cm^{-2}$ and $E_{53}= E/10^{53}\ergs$.

In Fig.~3 are shown the size and age of the supershell
as well as the volume number density of the ambient medium
with $E_{53}$ being a free parameter. 
For a range of input energy, $0.01\le E_{53}\le 100$,
we get the possible range of the other parameters $0.53\le n_1 \le 53$, 
$2\pc\le R \le 180\pc$, and $5\times10^2\yrs \le t \le 5\times 10^4 \yrs$.

From these values we estimate the total kinetic energy 
of the expanding supershell given by
$E_k=2\pi R^2 N_{HI} m_H v_{exp}^2= 7.3\times10^{52}\ E_{53} \ergs$.
It is also noticeable that this kinetic energy can be comparable 
to those of the galactic supershell including those in Our Galaxy, 
NGC 4631, and M101 (Heiles 1979, Rand \& van der Hulst 1993, Wang 1999).
Furthermore, the P-Cygni Ly$\alpha$ lines of the primeval galaxies 
also show the similar energy scale. Thus, we propose that the supershell
is the remnant of a hypernova or a GRB that had exploded earlier than
GRB971214.

\section{Summary and Implications}

We have studied on the formation of P-Cygni type Ly$\alpha$ 
in the spectrum of GRB971214, and found that there are at least
three components in the system, i.e. the parsec scale remnant 
of GRB971214 itself, a giant H II region, and a supershell surrounding it. 

The giant H II region from which Ly$\alpha$ emission originates 
is photoionized by a super stellar cluster whose total ionizing photons
correspond to those emitted from about $10^4$ O5 stars. 
The existence of the P-Cygni type absorption plausibly implies
that there exists a supershell surrounding the H II region.
By a profile fitting procedure, we found out that the shell is expanding 
with a velocity of $v_{exp}=1500\kms$ and its neutral column density 
$N_{HI}=10^{20}\cm^{-2}$. We also revised the redshift of the 
Ly$\alpha$ emission source to be $z=3.425$,
and its unobscured Ly$\alpha$ luminosity
to be $L_{Ly\alpha} = (1.8\pm0.8)\times 10^{42} \erg \s^{-1}$,
which gives a more reasonable star formation rate
to be $R_{SF} = (7\pm3)\ {\rm M}_\odot\yr^{-1}$.

We also applied the theory on the hydrodynamical evolution
of supernova remnants to the supershell surrounding GRB971214.
Assuming a reasonable scale of the initial explosion energy 
of the supershell, we propose 
that the supershell is at the adiabatic phase, 
with its radius $R = 18\ E_{53}^{1/2}\ \pc$, its age
$t = 4.7\times10^3\ E_{53}^{1/2}\ \yrs$, and the number density 
of the ambient medium $n_1 = 5.4\ E_{53}^{-1/2}\cm^{-3}$,
where $E_{53}= E/10^{53}\ergs$.
And we estimate the kinetic energy of the supershell to be
$E_k= 7.3\times10^{52} E_{53} \ergs$.

It is noticeable that there are many astronomical objects
showing the similar characteristics. Using the X-ray data, Wang (1999) has 
discovered the candidates of GRB remnants in M101, of which two exhibit 
similar physical characteristics to those of the GRB remnant in GRB971214.
With the advent of 10m class telescopes, a large number of primeval galaxies 
are observed by applying several methods including the Lyman break method 
(Steidel 1996).
About $50$ percent of Ly$\alpha$ emission lines in their spectra 
show P-Cygni type profiles. It is
suggested that these profiles are formed in an expanding media 
surrounding the star forming region and a case study was performed for
DLA~2233+131 in detail (Lee \& Ahn 1998).

Recently one of the most debated suggestions is the hypernova conjecture,
according to which gamma-ray bursts occur in star 
forming regions. Our results strongly favor this model and more
concrete evidence is expected as the sample of GRB spectra showing 
Ly$\alpha$ emission becomes statistically significant.

\acknowledgments
 
The author thanks George Djorgovski and Shri Kulkarni 
for kindly providing the optical spectrum of GRB971214.
He also thanks Bon-Chul Koo, Kee-Tae Kim, Hee-Won Lee, 
Hwang-Kyung Sung, In-Su Yi, and Hyung-Mok Lee for their 
invaluable discussions.
The author thanks to the anonymous referee for his/her
fruitful comments and suggestions.

\clearpage

\end{document}